\documentclass[a4paper,12pt]{article}
\usepackage{amssymb,xypic,latexsym,graphicx}
\usepackage{geometry}
\title{What `shape' is space-time?} 
\author{Timothy Porter\\ School of Informatics,\\University of Wales 
  Bangor,\\ Bangor,\\ Gwynedd, LL57 1UT, \\Wales, U.K.} 
 
\newtheorem{theorem}{Theorem} 
\newtheorem{proposition}{Proposition}

\begin{document} 
\maketitle 
   
\abstract{Some examples from the mathematics of shape are presented that question some of the almost hidden assumptions behind results on limiting behaviour of finitary approximations to space-time.  These are presented so as to focus attention on the observational problem of refinement and suggest the necessity for an alternative theory of `fractafolds' against which the limiting theory of $C^\infty$-differential manifolds usually underlying (quantum) general relativity can be measured.}

\section{ Introduction and Physical Background}

Even from the start, the space-time `manifold' was considered `unphysical'. It involved numerous powerful mathematical concepts that were inherently beyond observation, although providing apparently essential tools for developing the physical theory.  More recently this `unphysicality' has stimulated attempts to use an `observational' approach to model the differential, dynamic aspects of space-time (and eventually to quantise it) using discrete, algebraic or combinatorial models.

One of the finitary approaches to discrete space-time was pioneered by Sorkin, \cite{Sorkin}.  This approach assumes space-time is modelled by a manifold, $M$, then assumes an open cover ${\mathcal F}$ of $M$ is given.  This ${\mathcal F}$  is thought of as corresponding to the set of observations being considered.  If $ U \in {\mathcal F}$, then the events within $U$ are thought of as being operationally indistiguishable by that observation.  Of course, if $x \in U$ and $y  \notin U$, then the observation distinguishes $x$ and $y$.  If one considers the equivalence relation corresponding to `operational indistinguishability' relative to ${\cal F}$, the result is a $T_0$-space which is a `finitary substitute for $M$' with respect to the covering ${\cal F}$.

As has been remarked on elsewhere, e.g. in the papers of Raptis and Zapatrin, \cite{RapZap1,RapZap2}, this construction of Sorkin is closely related to that of the nerve of the open cover ${\cal F}$, a construction from algebraic topology usually attributed to \v{C}ech, \cite{Cech}.  This yields a simplicial complex that will be denoted $N(M,{\cal F})$ or $N({\cal F})$, if $M$ is understood.  The $T_0$ space given by Sorkin's construction is merely the abstract simplicial complex, $N(M,{\cal F})$, provided that $\cal F$ is \emph{irredundant} or \emph{minimal} (i.e. none of its elements can be omitted without loss of the property that it covers $M$) and \emph{generic} (that is, all non-empty intersections are different). (Without these two conditions, one still has that any $T_0$-space  has the homotopy type of a simplicial complex by an old result of McCord, \cite{McC}.)

In this `finitary substitute for $M$' game, a result of importance is that essentially all the important structure of $M$ can be recovered in the classical limit.  Although true, this seems to be a bit of `sleight of hand'.  It assumes $M$ is a manifold and retrieves $M$, but, ignoring questions of the physical interpretation of the limiting process beyond the Planck level, this still seems back to front.  To postulate $M$ is a manifold is unnecessary for a limit space to exist and the initial $M$ is known to be recoverable from the limit space. This suggests a slightly provocative pair of questions:
\begin{enumerate}
\item If $M$ was replaced with a more general space than a manifold, for instance a fractal version of a manifold (such `fractafolds' are known to exist, but have been little studied) how could the difference be observed?
\item If $M$ was replaced by a pointless space (i.e. a locale or better still a quantale (that is its non-commutative  `cousin')), how could one know the difference?
\end{enumerate}
In this paper we will concentrate on the first of these, not to answer it but to examine some toy examples to illustrate the problem, showing that it suggests some more thought should be given to the physical interpretation of the refinement of observations.  The examples are standard spatial examples from the mathematical area of Shape Theory and its better structured variant Strong Shape Theory.  This theory has strong links with parts of $C^*$-algebra theory (cf. Blackadar, \cite{Bla} and Dadarlat, \cite{Dad}).

\section{Finitary substitutes for topological spaces.}

The basic idea of Sorkin's `algorithm' is to replace the existing topology of a space $X$ by that generated by an open covering $\mathcal{F}$ of $X$. (To keep the examples simple, we will assume $X$ is a compact metric space, so that we can assume $\mathcal{F}$ is finite.)  This process is known as \emph{coarse graining} and is in many ways similar to the digitalisation process within image analysis, cf. \cite{MG}.  The idea of coarse graining is that `large', `coarse' or `fuzzy' open sets are nearer the reality of observation than are points, but that the geometric point-like underlying space $X$ will be recoverable at the ideal limit of infinite refinement of the $\mathcal{F}$s.

Formally Sorkin's construction is as follows:\\
Given $X$ and an open cover $\mathcal{F}$, introduce an equivalence relation $\equiv_\mathcal{F} $ on $X$ by:\\
\quad if $x, y \in X$,\\
\centerline{$x \equiv_\mathcal{F} y$ \quad if and only if \quad$\forall  U \in \mathcal{F}, x\in U \Leftrightarrow  y \in U$.}
The quotient $X_\mathcal{F} := X/\equiv_\mathcal{F} $ will be a finite topological space if $\mathcal{F} $ is a finite open cover.  We will give it the quotient topology with respect to the obvious projection
$$X \to X_\mathcal{F}. $$
This space $X_\mathcal{F}$ will be a $T_0$-space, and every point $x$ in $X_\mathcal{F}$ is in a unique minimal open set, $U_x$. This gives a partial order on $X_\mathcal{F}$ given by  
$$x \leq y \quad \Leftrightarrow\quad U_x \subseteq U_y \quad\Leftrightarrow\quad x \in U_y. $$

\vspace{1cm}

\textbf{Example 1.}

A circle, $S^1$, with open cover $\mathcal{F} = \{U_1, U_2, U_3\}$ with, in polar coordinates,
\begin{eqnarray*}U_1 &=& \Big(-\frac{2\pi}{3}, \frac{2\pi}{3}\Big);\\
                   U_2 &=& ``\Big(0, -\frac{2\pi}{3}\Big) " \mbox{\quad i.e.\quad } \Big(0, \pi\Big]\cup \Big(-\pi,-\frac{2\pi}{3}\Big);\\
                   U_3 &=& `` \Big( \frac{2\pi}{3},0\Big)" \mbox{ \quad i.e. \quad  } \Big(\frac{2\pi}{3},\pi\Big]\cup\Big(-\pi ,0\Big).
\end{eqnarray*}
Every point of $S^1$, with the exception of $0$, $\frac{2\pi}{3}$ and $ -\frac{2\pi}{3}$, is in exactly two of these.  There are, in all, six equivalence classes.  Picking three representatives for the non-singleton classes, we have the minimal open sets are as follows 
$$U_0 = U_{1}, \quad \quad  \quad \quad U_{\frac{2\pi}{3}} = U_2, \quad \quad \quad\quad U_{-\frac{2\pi}{3}}= U_3,$$
$$U_{\frac{\pi}{3}} = U_1\cap U_2 =:U_{12}, \quad U_{-\frac{\pi}{3}} = U_1\cap U_3 =:U_{13}, \quad U_{\pi} = U_2\cap U_3 =: U_{23}.$$
This gives a partially ordered set whose Hasse diagram is 
$$\xymatrix{
1\ar@{-}[d]\ar@{-}[drr]&2\ar@{-}[dl]\ar@{-}[d]&3\ar@{-}[dl]\ar@{-}[d] \\
12 & 23 & 13}
$$
Thus $S^1_\mathcal{F}$ has 6 points.  (The basic `geometry' of $S^1_\mathcal{F}$ is a circle and, in fact, its homotopy type is that of  $S^1$.)

\section{Nerves}

Another way of extracting information from a cover $\mathcal{F}$ of a space $X$ is by forming the nerve of the covering.  The construction gives a simplicial complex and can be attributed to \v{C}ech, \cite{Cech}.  First we recall the definition of simplicial complex.

A simplicial complex $K$ is a set of objects, $V(K)$, called \emph{vertices} and a set of finite subsets of $V(K)$, called \emph{simplices}.  The simplices satisfy the condition that if $\sigma \subset V(K)$ is a simplex and $\tau \subset \sigma$, then $\tau$ is also a simplex.

To each simplicial complex $K$, one can classically associate a topological space called the \emph{polyhedron} of $K$ (or sometimes the \emph{geometric realisation} of $K$) and denoted $|K|$.  This is formed from the set of all functions from $V(K)$ to $[0,1]$ such that
\begin{itemize}
\item if $\alpha \in |K|$, the set 
$$\{v \in V(K) ~|~ \alpha(v)  \not=  0\}$$
is a simplex of $K$;
\item \hspace{2cm} $\displaystyle \sum_{\alpha \in V(K)} \alpha (v) = 1$.
\end{itemize}
If $s\in K$ we denote by $|s|$ the set $$|s| = \{\alpha \in |K| ~\Big|~ \alpha(v) \neq 0 \mbox{ implies } v \in s\},$$
and $$\langle s \rangle = \{ \alpha \in |K| ~\Big|~ \alpha(v) \neq 0  \mbox{ if and only if } v \in s\}.$$
$\alpha (v)$ is called the $v^{th}$ barycentric coordinate of $\alpha$ and the mapping from $|K|$ to $[0,1]$ defined by $p_v(\alpha) = \alpha(v)$ is called the $v^{th}$ barycentric projection.

We can put a metric $d$ on $|K|$ by $$d(\alpha,\beta) = \Big(\sum_{v\in V(K)} (p_v(\alpha) - p_v(\beta))^2\Big)^\frac{1}{2}$$and the topology that results is the initial topology for the barycentric projections. (For more on $|K|$, see Cordier and Porter, \cite{JMCTP}, Ch. 3 or most books on algebraic topology, e.g. Spanier \cite{Spa}.  Recently non-commutative geometric realisations for simplicial complexes have been introduced by Cuntz and his collaborators \cite{Cuntz}, but we will not be looking in that direction in this paper.  Other realisations can also be useful, cf. the discussion in \cite{MG}.)

\medskip

\textbf{From coverings to simplicial complexes.}

As before, let $X$ be a space and $\mathcal{F}$ be an open covering of $X$.  The nerve of $\mathcal{F}$, denoted $N(X,\mathcal{F})$ or $N(\mathcal{F})$ if no confusion will arise, is the simplicial complex having as vertices the open sets in $\mathcal{F}$ and for simplices those finite families of open sets in $\mathcal{F}$ whose intersection is non-empty:

\centerline{ i.e. $\{U_0, \ldots, U_n\} \subset \mathcal{F}$ is a simplex of $N(\mathcal{F})$ if and only if $\displaystyle \bigcap^n_{i=0}U_i \neq \emptyset.$}
(This would be an $n$-simplex and we will write $\langle U_0, \ldots, U_n\rangle$ for it.)

\medskip

\textbf{Example 1 revisited.}

For $X = S^1$, $\mathcal{F} = \{U_1, U_2, U_3\}$ as before,\\
\hspace*{1cm} Vertices of $N(\mathcal{F}) = \langle U_1\rangle, \langle U_2\rangle, \langle U_3\rangle$,\\
\hspace*{1cm} 1-simplices of $N(\mathcal{F}) =\langle U_1, U_2  \rangle, \langle U_1, U_3  \rangle, \langle U_2, U_3 \rangle$.\\
Schematically one might represent $N(\mathcal{F})$ by a diagram\\
$$\xymatrix{&2\ar@{-}[dl]_{1,2}\ar@{-}[dr]^{2.3}&\\
1\ar@{-}[rr]_{1,3}&&3}$$

\medskip

Any simplicial complex determines a poset (partially ordered set) by subset inclusion of the simplices.  It is clear in this example that the resulting poset is the opposite of that representing $X_\mathcal{F}$.

\medskip

Various remarks are in order here.  Firstly, this is standard elementary material from algebraic topology, so we are, in some sense, not doing that much as yet, however the physical interpretation of $\mathcal{F}$ implies various physical interpretations for constructions based on $\mathcal{F}$.  For instance, a point in $|N(\mathcal{F})|$ corresponds to a function $\alpha$ from the `set of observations', $\mathcal{F}$ to $[0,1]$ and the conditions imply it is a linear (in fact, convex) combination of the observations, so interpolates  between them.  The coordinates of  points of $|N(\mathcal{F})|$ can also be thought of as measuring `levels of confidence', `levels of contribution' or a sort of `fuzzy combination' of the attributes corresponding to the observations.  We may sometimes use the term `fuzzy superposition of observations' to try to capture this intuition. 

The close fit between the poset associated with $|N(\mathcal{F})|$ and that of $X_\mathcal{F}$ is due to the fact that the covering $\mathcal{F}$ in the example is minimal and generic (as mentioned earlier).  More precisely:

A covering $\mathcal{F}$ is \emph{minimal} or \emph{irredundant} if for any $U \in \mathcal{F}$, $\bigcup(\mathcal{F}\setminus \{U\}) \neq X$, i.e. none of the open sets in the covering can be left out without missing out some of $X$ from the region covered.

The covering $\mathcal{F}$ is \emph{generic} if all non-empty intersections are distinct.  If $\mathcal{F}$ is generic, an intersection $A_0 \cap \ldots \cap A_n$ of sets in $\mathcal{F}$ is \emph{uniquely} determined by the simplex $\langle A_0, \ldots, A_n\rangle$ of $N(\mathcal{F})$.

Assume that $\mathcal{F}$ is minimal and generic.

With any equivalence class $[x]_\mathcal{F}$ of points under the equivalence relation $\equiv_\mathcal{F}$, we can associate the simplex
$$P_x = \{ A \in \mathcal{F} ~\big|~ [x]_\mathcal{F} \subset A\}.$$
Of course, by definition
$$x \equiv_\mathcal{F} y \Leftrightarrow \forall  U \in \mathcal{F},( x\in U \Leftrightarrow y \in U).$$
The point $[x]_\mathcal{F}$ of $X_\mathcal{F}$ determines a minimal open set of $X_\mathcal{F}$  as there are only finitely many open sets in $X_\mathcal{F}$.  Explicitly the minimal open set $U_x$ containing $[x]_\mathcal{F}$ is the intersection of the open sets of $X_\mathcal{F}$  that contain $[x]_\mathcal{F}$.  As the space $X_\mathcal{F}$  has the quotient topology with respect to the projection 
$$p : X \to X_\mathcal{F},$$ 
we have
$$p^{-1}(U_x) = \{p^{-1}(U) \big| [x]_\mathcal{F} \in U\} = \bigcap P_x.$$
As $\mathcal{F}$ is generic, $[x]_\mathcal{F}$ determines $P_x$ and vice versa.  The simplexes of $N(\mathcal{F})$ are ordered by the face relation, or equivalently by inclusion.  The imnplication of the minimality of $\mathcal{F}$ is that each set $U$ of $\mathcal{F}$ is needed for at least one point, i.e. there is an $x\in U$ that is in no other set of the covering, hence $p^{-1}(U_x) = U$ and thus the vertices of $N(\mathcal{F})$ corrrespond to points in $X_\mathcal{F}$ .  Without minimality, there could be vertices of $N(\mathcal{F})$ that do not correspond to minmal elements of the poset of the topology of $X_\mathcal{F}$.  (As an example take the cover $\mathcal{F} = \{ U_1, U_2, U_3\}$ of the circle, $S^1$, in example 1.  Form a non-minimal covering by adding in $S^1$ itself into the covering giving $\mathcal{F}_1 = \{ U_1, U_2, U_3,S^1\}$.  There is no effect on $X_\mathcal{F}$, but $N(\mathcal{F}_1)$ takes the form of a cone with base $N(\mathcal{F})$.  It has one extra vertex $\langle S^1 \rangle$, three extra 1-simplices $\langle U_i, S^1\rangle$, $i = 1,2,3$, and has 2-simplices as well, $\langle U_1, U_2, S^1\rangle$, etc.)

We thus have that the poset of $X_\mathcal{F}$ and the face poset of $N(\mathcal{F})$ have ``the same'' points. The correspondence, of course, reverses the order on the points:
$$x \leq y \Leftrightarrow  U_x \subseteq U_y,$$
but if $\tau \subset \sigma$ in $N(\mathcal{F})$, then $\bigcap \tau \supset \bigcap \sigma$.

\medskip

The face poset of any simplicial complex, $K$, itself yields a new simplicial complex.  If $P$ is any poset, we form its nerve $N(P)$ (with a slightly different meaning of `nerve') by taking the points of $P$ as vertices of $N(P)$ and taking as simplices the totally ordered finite subsets of $P$, i.e. the chains:
$$\sigma = \{ p_0 < p_1 < \ldots < p_q\} \in N(P)_q.$$
If we apply this construction to a face poset, we get an abstract simplicial complex corresponding to an important subdivision of $K$.  As this `barycentric subdivision' will be needed when discussing triangulations later \underline{and} there is a physical interpretation of its vertices, we will recall its main attributes.  First we need the exact sense in which `subdivision' will be used.

If $K$ is a simplicial complex, a \emph{subdivision} of $K$ is a simplicial complex, $K^\prime$, such that\\
a) the vertices of $K^\prime$ are (identified with)  points of $|K|$;\\
b) if $s^\prime$ is a simplex of $K^\prime$ , there is a simplex, $s$ of $K$ such that $s^\prime \subset |s|$; and\\
c) the mapping from $|K^\prime|$ to $|K|$, that extends the mapping of vertices of $K^\prime$ to the corresponding points of $|K|$, is a homeomorphism (thus continuous with a continuous inverse).

\medskip

Our earlier interpretation of the points of $|K|$ as combinations of the `observations' corresponding to the vertices, allows a physical interpretation to be ascribed to $K^\prime$.  The barycentric subdivision is one of the best known and most useful natural subdivisions available in general.  (Other are also used, for instance, the middle edge subdivision.)  The barycentric subdivision has the good property that it exists without recourse to the realisation process, although usually introduced via that process.

If $\sigma = \{ v_0, \ldots, v_q\} \in K_q$, the set of $q$-simplices of a simplicial complex, $K$, then its barycentre, $b(\sigma)$, is the point $$b(\sigma) = \sum_{0\leq i \leq q}\frac{1}{q + 1} v_i \in |K|.$$
The barycentric subdivision $sd K$ of $K$ is the simplicial complex whose vertices are the barycentres of the simplices of $K$ and whose simplices are finite non-empty collections of barycentres of simplices, which are totally ordered by  the face relation of $K$.

As $b(\sigma)$ is completely determined by $\sigma$, this can be rephrased as:
\begin{itemize}
\item the vertices of $sd K$ \emph{are} the simplices of $K$;
\item the subset $\{\sigma_0, \ldots \sigma_q\}$ is a simplex of $sd K$ if there is a total order on its elements (which will be assumed to be the given order, as written) such that
$$\sigma_0\subset  \ldots \subset\sigma_q.$$
\end{itemize}

For any open cover $\mathcal{F}$ ( so system of `observations') and any  $U_0, \ldots, U_q$ with $\bigcap U_i \neq \emptyset$, the barycentre of $\sigma = \langle U_0, \ldots, U_q\rangle,$ is the `averaged' observation of the simplex, alternatively, it is the renormalised union of the observations.  Thus one can attempt to assign a physical meaning to the vertices of $sd N(\mathcal{F})$.  Comparison  with the general construction, $N(P)$, for the case where $P$ is the face poset of $N(\mathcal{F})$ shows that $sd N(\mathcal{F})$ is just the nerve of the face poset of $N(\mathcal{F})$, and so we are back with $X_\mathcal{F}$, but viewed slightly differently.  The important point to note is that the nerve $N(\mathcal{F})$ and thus its barycentric subdivision, \emph{does not require the unphysical space $X$ to consist of points for it itself to make sense}.  Any space determines a locale / topology which is a lattice of the open sets of the space.  Within that context, `covering' makes perfect sense as does the nerve of such a `covering' and hence $X_\mathcal{F}$ is a valid construct even if $X$ has no `points', (cf. the general use of `pointless topology' by Isham et al. \cite{IshBut}).

\section{Triangulations}

In many papers on quantum gravity, the structures are imposed on a `triangulated manifold'.  This, in turn, leads to a dual decomposition of the manifold to which are assigned representations of some group or perhaps, quantum group.  For us, triangulated spaces provide a good source of examples to use to investigate what is the observational and informational content of the nerves of open covers.  We will use a classical definition and some associated results (cf. Spanier, \cite{Spa}, p. 114 and p. 125).

A \emph{triangulation} $(K,f)$ of a space $X$ consists of a simplicial complex $K$ and a homeomorphism $f :  |K|\to X$.  We will usually confuse $|K|$ with $X$ and call $X$ a \emph{polyhedron} in this case.  Given any vertex $v$ of $K$, its \emph{star} is defined by 
$$st( v) = \{ \alpha \in |K|~ \Big| ~\alpha(v)\neq 0\}.$$
The set, $st(v)$, is open in $|K|$ and, if $\alpha : V(K) \to [0,1]$ is loosely interpreted as a `fuzzy observation' or measure of a `confidence level', then  $st(v)$ consists of those such `fuzzy observations' that `observe' the notional point, $v$.  We have 
$$st(v) = \bigcup \{\langle s \rangle ~\big| ~ v \mbox{ is a vertex of }s\},$$
the union of the interiors of those simplices that have $s$ as a vertex.  These vertex stars give an open covering of $|K|$ and the following classical result tells us that the nerve of this covering is $K$ itself (up to isomorphism):

\begin{proposition}(cf. Spanier \cite{Spa}, p. 114)

Let $\mathcal{U} = \{st(v) ~\big|~ v \in K\}$.  The vertex map $\phi$ from $K$ to $N(\mathcal{U})$ defined by 
$$\phi (v) = \langle st(v)\rangle$$
is a simplicial isomorphism
$$\phi : K \cong N(\mathcal{U}).$$\hfill $\blacksquare$
\end{proposition}
 As an example, the triangle, as simplicial complex, has vertices
$$V(K) = \{ 1, 2, 3\} $$
and simplices $\{1\}, \{2\}, \{3\}, \{1,2\}, \{1,3\}, \{2,3\}$.  (This is the \underline{triangle} not the \underline{2-simplex}, so there is no 2-dimensional face.)  This obviously provides a triangulation of the circle, $S^1$, and this can be done in such a way that the vertex star covering of $S^1$ that results is precisely that considered in example 1.
\medskip

The above result, and the example, illustrate that for polyhedra (and thus for triangulated manifolds), the approach via triangulations is at least as strong as that via open coverings.  Another classical result (Spanier, \cite{Spa}, p.125) tells us that they are of equal strength:
\begin{theorem}

Let $\mathcal{F}$ be any open covering of a (compact) polyhedron $X$.  Then $X$ has triangulations finer than $\mathcal{F}$  \hfill $\blacksquare$
\end{theorem}

Here a triangulation $(K,f)$ of $X$ is said to be \emph{finer than} $\mathcal{F}$  if for every vertex $v \in K$, there is a $U \in \mathcal{F}$ such that $f(st(v)) \subset U$. Note that compactness is not necessary for the result, but if $X$ is compact, the triangulations can be obtained via iterated barycentric subdivisions.

Thus, for polyhedra, the triangulations give open coverings and, modulo refinement of coverings, we can always replace an open covering of a polyhedron by a vertex star covering. This has two implications for us.

(i) Given the finitary substitutes of a polyhedral space, $X$, with respect to vertex star coverings, we clearly will get the space back again in the limit.  Thus the results on retrieval of the underlying `manifold' of spacetime from the finitary substitutes, although new results mathematically, were not that surprising.  If one tests for possible non-manifold models of space-time, then each finitary substitute gives a polyhedral space whose physical significance may be examined via the realisation of the poset.   The key to its `non-manifoldness' must lie not in the finitary substitutes themselves, as they would yield polyhedra, but in the relationships \emph{between} the finitary substitutes for different coverings, that is the \emph{refinement maps}.

(ii) These results say that for polyhedral spaces, open coverings can always be replaced by vertex star coverings that are `finer'. This focusses attention on the comparison between triangulations as one aspect of the problem of refinements mentioned in (i). This is easier to see in the case of polyhedra than is the general refinement map problem for general open coverings of general spaces.

\medskip

There is thus here an interaction between several concepts.  Basing models on a `manifold', there are a large number of theories that use triangulations of that manifold to define the physical model (cf. all the material on spin networks, spin foams, etc. and much of quantum field theory in general).  There is some good physical intuition underlying these triangulations, but if one bases the triangulation on `observations', as in the combinatorial  interpretation of Sorkin's theory, it would seem that the triangulation is not `imposed' on the `manifold', a terminology sometimes used leading to questions of `invariance' of some theory and `independence' of an `invariant' from the choice of the `imposed triangulation'.  If the Sorkin model is used, the triangulation is not `imposed', it is `observed' or rather \emph{it organises the observations}!  This change of perspective is fairly benign, but in some calculations, e.g. within spin network theory, the number of top dimensional simplices is allowed to go to infinity as part of the construction (this is usually obtained by taking the dual of the triangulation that gives a cell complex structure to the manifold, and then this  number is the number of vertices of that cell complex.)  If the underlying space (or space-time) is a manifold, the limiting process is fairly uncontentious, although, typically, questions of convergence of certain power series, or path-integral analogues, do cause problems, but the assumption that there is an underlying `manifold' itself is all important here, yet receives scant attention.  What if it was not a manifold?  What if it was not even a polyhedron?   (The author believes these are not the right questions, but, even so, feels that to get a perspective on what are the right ones, they are a possible first step.)  In the next section, some `toy' non-manifolds will be `observed' and compared with some polyhedra.

\section{Refinements: what can go wrong?}

If the `triangulations' are thought of as the netwoeks of interacting observables, one has to ask how to compare two such networks.  To start with assume that we have an underlying simplicial cp,[lex $K$ (or if you prefer the corresponding polyhedron, $|K|$).  To compare simplicial complexes, one uses `simplicial maps':

If $\phi : K_1 \to K_2$ is a simplicial map, then $\phi$ is a mapping from the vertices of $K_1$ to those of $K_2$ such that if $s$ is a simplex of $K_1$ (so $s \subset V(K_1)$), then $\phi(s)$ is a simplex of $K_2$.

\medskip

\textbf{Example}

Consider $K_2$ to be the triangle (with $V(K_2) = \{1,2,3\}$) and $K_1 = sd K_2$, its barycentric subdivision.  Intuitively one would expect a `sensible' simplicial mapping from $K_1$ to $K_2$ and that there would be some `natural' choice of one.  Recall that the vertices of $K_1$ are $\{1\}$, $\{2\}$, $\{3\}$, $\{1,2\}$, $\{1,3\}$, $\{2,3\}$.  If the mapping $\phi$ is to be as `natural' as possible then one would expect $\phi(\{1\}) = 1$, etc., but what should be the image of $\{1,2\}$.  It can be either $\{1\}$ or $\{2\}$.  If it is $\{1\}$, then the 1-simplex $\{\{1\},\{1,2\}\}$ gets squashed to the vertex $\{1\}$, whilst $\{\{2\},\{1,2\}\}$ goes to $\{1,2\} \in K_2$.  Similarly if $\phi(\{1,2\}) = 2$, $\phi$ will squash $\{\{2\},\{1,2\}\}$  and `expand' $\{\{1\},\{1,2\}\}$ .  There will, of course, be similar choices to be made for $\{1,3\}$ and $\{2,3\}$.

\medskip

\textbf{Remark}

It is instructive to examine what these approximations to the identity map on the triangle look like if considered via the Sorkin model:\\
there is absolutely no choice when defining the map. More exactly the triangulation of the triangle and its barycentric subdivision yield two vertes star open coverings of $S^1$.  For the coarser one, one gets the poset of example 1:
$$\xymatrix{\ar@{}[d]_{{X_\mathcal{F}}}  &1\ar@{-}[d]\ar@{-}[drr]&2\ar@{-}[dl]\ar@{-}[d]&3\ar@{-}[dl]\ar@{-}[d] \\
&1,2 & 2,3 & 1,3}
$$
and the barycentric subdivision (with, it is hoped, the obvious notation):
$$\xymatrix{\ar@{}[d]_{{X_{\mathcal{F}^\prime}}}  &1\ar@{-}[d]\ar@{-}[drrrrr]&12\ar@{-}[dl]\ar@{-}[d]&2\ar@{-}[dl]\ar@{-}[d]&23\ar@{-}[dl]\ar@{-}[d]&3\ar@{-}[dl]\ar@{-}[d]&13\ar@{-}[dl]\ar@{-}[d] \\
&112 & 212 & 223&323&313&113} 
$$
The finitary approximation to the identity yields a map of posets $$X_{\mathcal{F}^\prime} \to X_\mathcal{F},$$
which collapses the links $112\leq 12$, $212 \leq 12$, etc., to identities, so every other vertex in the top row gets sent to a vertex in the second row of $X_\mathcal{F}$.

This corresponds at the level of the nerves of these posets to the fact that there \underline{is} a natural map from the double barycentric subdivision of the triangle to $K_1$, namely that  which collapses the middle two edges of each edge, ($a$ and $b$ in the diagram,
$$\xymatrix{v_1  \ar@{-}[r]& \bullet \ar@{-}[r]^{a}&v_{12} \ar@{-}[r]^b&\bullet \ar@{-}[r]&v_2,)}$$ to $v_{12}$, (cf. Landi et al, \cite{landietal}).

This looks simple but unfortunately its simplicity is due largely to the one dimensionality of the example.

\medskip

The point of the example is that at the level of simplicial complexes, refinement yields non-determinism.  There is no natural choice of simplicial map.  At the level of the Sorkin model, there is no such problem, but if the physicality of the underlying space, based on non-physical points, is questioned, then there is a second question relating to the atomicity of observations.  If open sets correspond to observations in the physical semantics of the mathematics, and intersections correspond to combined observations, then which are the `atomic' observations and which the combined ones?  Mathematically, which of the vertices of the barycentric subdicvision are original ones and which are ones added as barycentres of `original' simplices. (Remember the barycentres are averages of the surrounding vertices/ observations).  In the end, this problem may be not that difficult to handle, but it is the physical analogue of a mathematically challenging problem, that of the possiblity of choosing refinements in a coherent way.  It is to this problem we will need to turn shortly.

If we make no assumptions about the space $X$ which will be `observed', i.e. we do not assume given a triangulation nor that it is a polyhedron, there is still an obvious definition of a refinement of an open covering, and thus of refined  observations. Again the definition is `classical'.

Suppose $\mathcal{U}$ and $\mathcal{V}$ are two open coverings of $X$, then $\mathcal{V}$ is a \emph{refinement} of $\mathcal{U}$ if for any $V \in \mathcal{V}$, there is some $U \in \mathcal{U}$ with $V\subseteq U$.  A \emph{refinement map} (called by Spanier, \cite{Spa}, p.152, a canonical projection) is a function
$$\varphi :\mathcal{V} \to \mathcal{U}$$on the indexing sets of the open coverings such that for any $V \in \mathcal{V}$ $$V\subseteq \varphi(V).$$

\textbf{Example}

If $\mathcal{U}$ and $\mathcal{V}$ are vertex star coverings corresponding to a triangulation of a polyhedron and its barycentric subdivision, then refinement maps were discussed in the previous example.

\medskip

All a refinement map does is represent the refinement relation by a definite choice.  Different choices are related.

If $\varphi :\mathcal{V} \to \mathcal{U}$ is a refinement map, then at the level of nerves, there is an induced simplicial map
$$N(\varphi) :N(\mathcal{V}) \to N(\mathcal{U})$$
mapping $\langle V_0, \ldots, V_p\rangle$ to the simplex determined by the set $\{\varphi(V_i), i = 0, \ldots, p\}$.  (Note, repetitions \emph{can} occur in the `list' $\varphi( V_0), \ldots,\varphi( V_p)$, so within the context of simplicial complexes that list may not be a simplex itself, but will determine one as $$\emptyset \neq \bigcap V_i \subseteq \bigcap \varphi(V_i).$$This slightly annoying technicality can be avoided by using simplicial sets rather than simplicial complexes, but this would require too big a detour here, so the difficulty will be `swept under the carpet', i.e. ignored!)

\textbf{Remark}

If $\varphi_0$, $\varphi_1$ are two refinement maps from $\mathcal{V}$ to $\mathcal{U}$, then $N(\varphi_0)$ and $N(\varphi_1)$ are `contiguous' simplicial mappings and are thus homotopic.  This suggests that the informational content of the inverse system of nerves, indexed by open coverings and refinement maps, may be filtering out details that correspond to the choices made by the refinement process.  Again this problem with the refinement process may need examination at the physical/ philosophical level, but none of the sources on the physical interpretation of the nerves has yet addressed the problem explicitly, partially since the problem seems to be hidden in the Sorkin theory.

The main point of refinement maps is that the induced simplicial maps between the nerves seem to encode subtle information on the underlying spaces.  The examples, below, in part explore this encoding:
\medskip

\textbf{Example 2: Warsaw circle}

The space $X$ may have very little separating it from `manifoldness'. yet a `singularity' can cause havoc! (The example is known as the Warsaw Circle as it was studied extensively by K. Borsuk and his Polish collaborators, cf. Borsuk,  \cite{Bor}.)

The Warsaw circle $S_W $ is the subset of the plane, $\mathbb{R}^2$, specified by
$$\{(x,\sin(\frac{1}{x}) ~|~ 0 < x \leq \frac{1}{2\pi}\}\cup \{(0,y) ~|~ -1 \leq y \leq 1\} \cup C,$$ where $C$ is an arc in $\mathbb{R}^2$ joining $(0,0)$ and $(\frac{1}{2\pi}, 0)$, disjoint fom the other two subsets specified above except at its endpoints.

\centerline{\includegraphics[width=3cm]{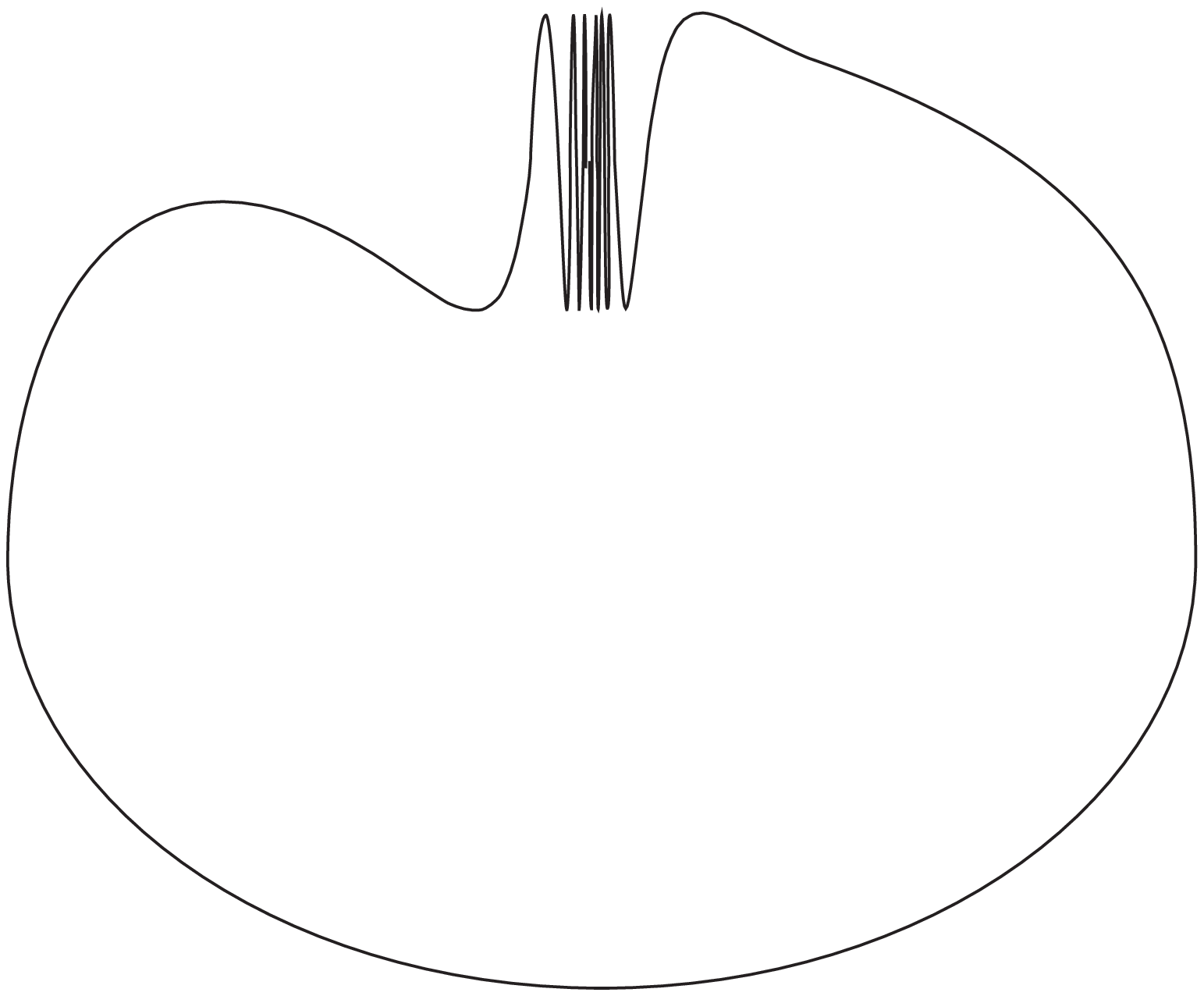}}

If one considers an open neighbourhood of $S_W$ in $\mathbb{R}^2$, say all points within $\frac{1}{n}$ of the set, then it looks like an annulus with a thickenning at one section.  Any open covering of $S_W$ by small open balls will have a nerve that is essentially the same as this, i.e. a circle with a thickened `bar' at one point, transverse to the circle. (Note that the interval on the $y$-axis is included to make the space compact.)
The Sorkin finitary substitute will, of course, be very similar.  Neither method can really tell the difference between $S_W$ and a space obtained by adding in a thin rectangle transverse to a circle at one point:

\centerline{\includegraphics[width=3cm]{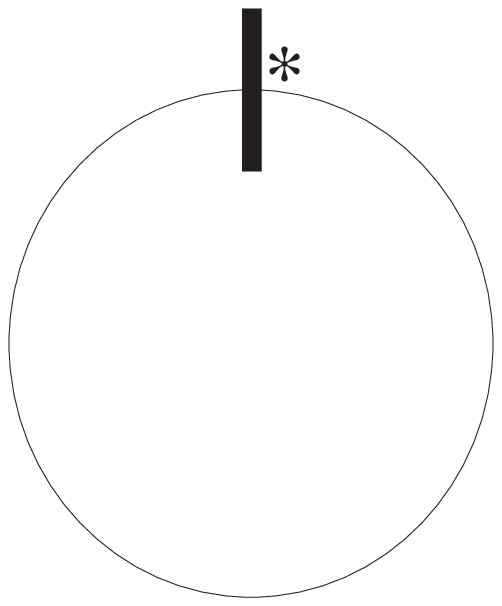}}

 For different open coverings, the only difference will be where the point of attachment (marked $\ast$) will occur and the thinnes of the rectangle.  The line of singularities given by the interval $[-1,1]$ on the $y$-axis cannot be observed, of course.  If one passes to finer and finer covers, most of the curve does not change appreciably, but the part near $\ast$ will `spawn' a very large number of vertices in both the nerve and the Sorkin models.

Within the htoery of Shape (cf. Borsuk, \cite{Bor}, Cordier-Porter, \cite{JMCTP}, Marde\v{s}i\'{c}-Segal,\cite{MarSeg}) or its stronger form (cf. Marde\v{s}i\'{c}, \cite{Mar}), this space $S_W$ has the same shape as $S^1$.  The direct physical interpretation of shape theory has yet to be explored.  That it has one is probable as its dual manifestation within $C^\ast$-algebra theory seems closely linked to deep properties of non-commutative geometry (cf. Blackadar, \cite{Bla}, Dadarlat, \cite{Dad} and Anderson and Grodal, \cite{AndGro}).  Here the question raised by this example is simply, how could a space time modelled by spaces that contained singularities of such a form as this be distiguished from a manifold model or more precisely, a polyhedral model.

The 1-dimensionality of the example is no hindrance.  It is easy to construct models of 2, 3, 4 or higher dimensions containing higher dimensional analogues of this singularity.

\textbf{Remark:}

A similarity exists between the above and methods of numerical analysis.  In the finite element method, for instance, it is often necessary to decrease the mesh size near singularities to ensure that the approximation is a sensible one there.  Perhaps a closer comparison of both the theories and the underlying intuitions in that situation and ours might shed light on the problem of refinement.  (The solution of the differential equation is the underlying `reality' and the numerical method `observes' that `reality', interpolating the information by an averaging process.)

\medskip

In example 2, the refinement maps are (in the limit) homotopy equivalences whether viewed in the \v{C}ech nerve or the Sorkin model.  The next example shows that this is often not the case.  Again it is a classical example from shape theory.

\textbf{Example 3.}

Let $A$ denote the solid torus obtained by rotating about the $z$-axis in $\mathbb{R}^3$, the disc centre $(0,2,0)$ of radius 1.  If $k$ is some natural number greater than 1, let $C_k$ be the curve in $A$
$$((2+\frac{1}{2}\sin u)\cos ku, (2 + \frac{1}{2}\sin u)\sin ku, \frac{1}{3} \cos ku)$$
for $u \in [0,2\pi]$.  Then there is an $\varepsilon_k > 0$ such that 
$$A_k = \{x : \rho(c, C_k) \leq \varepsilon_k\}$$
is contained in the interior of $A$ and is homeomorphic to $A$.  Let $$h_k : A \to A_k$$be that homeomorphism and $\overline{h}_k : A \to A$ the composite of $h_k$ with the inclusion.

Now suppose $k_1, k_2, \ldots $ is a sequence of integers $\geq 2$ and set $k_0 = 1$.  Denote by $g_0$ the identity map on $A$ and set 
$$g_m  = \overline{h}_{k_1}\overline{h}_{k_2}\ldots\overline{h}_{k_m} : A \to A$$
and let $B_m$ be the image of $g_m$, then $B_{m+1}$ is in the interior of $B_m$ and $B_m$ is homeomorphic to $A$.

To make this more accessible consider the case $k_1 =  k_2 = \ldots =  2$.  Then $C_2$ is a curve within the solid torus going twice around the hole.  Thus it looks something like this:

\centerline{\includegraphics[width=5cm]{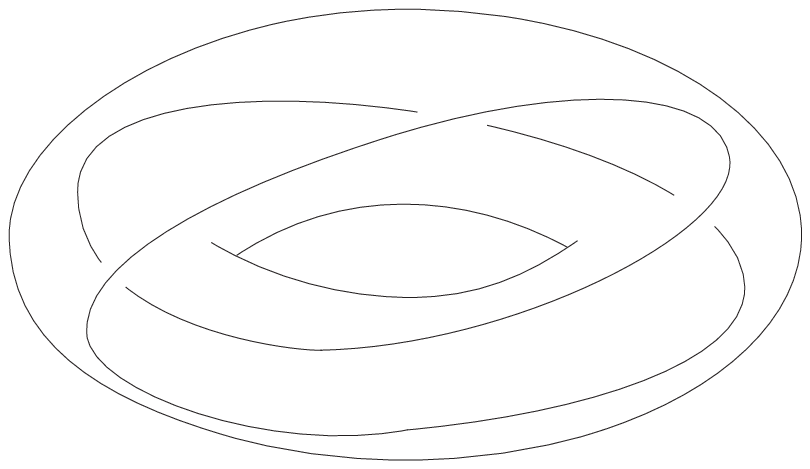}}

Thickenning this up, but not too much, so it gives an embedded solid torus in $A$, gives us $B_1$.  Taking this $B_1$ within $A$, map $A$ homeomorphically onto $B_1$.  The smaller solid torus is mapped to a new one, within $B_1$, going 4 times around the central hole of $A$.  This will be $B_2$, now continue.

Returning to  the general case, let $\underline{k} = (k_1, \ldots , \ldots)$ be the sequence and set 
$$S(\underline{k}) = \bigcap B_m.$$
This is a non-empty continuum called the \emph{$\underline{k}$-adic solenoid}.  The special case when all the $k_m$ are 2 is called the \emph{dyadic solenoid}.  Here is not the place to discuss the properties of these solenoids in detail, but two points need noting.

If we let $\kappa_m :S^1 \to S^1$ be the self map of the circle defined by 
$$\kappa_m(e^{i\theta}) = e^{ik_m\theta},$$
then writing $\underline{k} = \{k_m\}$ for the sequence as before, the $\underline{k}$-adic solenoid
$$S(\underline{k}) \cong \lim\{ \ldots \to S^1 \stackrel{\kappa_m}{\to} S^1\to \ldots \stackrel{\kappa_1}{\to} S^1\},$$
the inverse limit of circles with the $\kappa_i$ as structure maps.

The solenoids occur as natural examples of fractals and of strange attractors in dynamical system theory.  If one picks any open cover of $S(\underline{k})$, it can be refined to one of which the nerve is a circle.  Thus also the corresponding Sorkin finitary approximation is basically a circle as well.  To our observers, it will seem to be a circle at every refinement, yet the refinement mappings will not be homeomorphisms or even homotopy  equivalences - far from it.  For $S(\underline{2})$, the dyadic solenoid, they are composites of degree 2 mappings, i.e. double coverings if you prefer.  It is this fact that makes $S(\underline{2})$ or $S(\underline{k})$ in general, very singular (in fact, totally disconnected).

There are few, if any, non-constant continuous mappings between solenoids $S(\underline{k})$ and $S(\underline{\ell})$ for sequences $\underline{k}$ and $\underline{\ell}$ that are very different.  There are no maps from $S^1$ into any $S(\underline{k})$ other than trivial constant ones, although from $S(\underline{k})$ there are lots of maps to $S^1$.  These spaces are very far from being manifolds, yet all `observations' will see them as such.

\medskip

Here are some na\"{\i}ve questions to finish with.  Could space-time be fractal?  Could we find out if it was?  If the universe we live in, with the fundamental constants having the values they have, is optimal for some yet unknown set of conditions, might the result be a (strange) attractor in some larger systems of potential universes.  What could be the conditions on such an attractor that would be consistent with our observations both of its manifold-like nature (that is adequate for many purposes) and its granularity at the quantum scale?  \emph{What on earth does refinement of observations really tell you?}

\section{Conclusion.}

If one puts aside the idea that space-time is a manifold, then one can, of course, still apply the Sorkin finitary approximation `algorithm' to get useful information.  Using the observed connection with the \v{C}ech nerve construction and barycentric subdivision, the insights of shape theory suggest that further thought needs to be applied to the question of refining observations.  In particular it provides examples of spaces which observationally could be thought of as polyhedra, but are far from it.  These `toy models' have singularities of various kinds, some requiring finer and finer observations near a line of singularities, others everywhere singular.  The problem of refinement seems to have received very little attention in the now considerable physics literature on this area, but could clearly benefit from  a more detailed analysis.

\end{document}